\documentclass[12pt]{iopart}

\usepackage{iopams}
\usepackage{hyperref}
\hypersetup{colorlinks=true, urlcolor=red, citecolor=blue, linkcolor=blue}
\usepackage{graphicx}
\usepackage{subfigure}
\usepackage{epstopdf}
\usepackage{epsfig}

\begin{document}

\renewcommand*{\DefineNamedColor}[4]{%
   \textcolor[named]{#2}{\rule{7mm}{7mm}}\quad
  \texttt{#2}\strut\\}

\definecolor{red}{rgb}{1,0,0}
\definecolor{cyan}{cmyk}{1,0,0,0}

\title[Nonclassical properties...]{Nonclassical properties of states engineered by superpositions of quantum operations on classical states}

\author{Arpita Chatterjee\(^1\), Himadri Shekhar Dhar\(^1\) and Rupamanjari Ghosh\(^{1,2}\)}

\address{\(^1\)School of Physical Sciences, Jawaharlal Nehru University, New Delhi 110067, India\\
\(^2\)School of Natural Sciences, Shiv Nadar University, Gautam Budh Nagar, UP 203207, India
}
\ead{rghosh.jnu@gmail.com}

\begin{abstract}
We consider an experimentally realizable scheme for manipulating quantum states using a general superposition of products of field annihilation ($\hat{a}$) and creation ($\hat{a}^\dag$) operators of the type ($s \hat{a}\hat{a}^\dag+ t \hat{a}^\dag \hat{a}$), with $s^2 + t^2 = 1$. Such an operation, when applied on states with classical features, is shown to introduce strong nonclassicality. We quantify the generated degree of nonclassicality by the negative volume of Wigner distribution in the phase space and investigate two other observable nonclassical features, sub-Poissonian statistics and squeezing. We find that the operation introduces negativity in the Wigner distribution of an input coherent state and changes the Gaussianity of an input thermal state. This provides the possibility of engineering quantum states with specific nonclassical features.
\end{abstract}

(Some figures in this article are in colour only in the electronic version)

\pacs{42.50.Dv, 42.50.Ct}

\noindent{\it Keywords}: quantum state engineering, nonclassical states, volume of negativity, sub-Poissonian statistics, squeezing

\submitto{\JPB}
\maketitle

\section{Introduction}
\label{sec1}

In recent years, the generation and manipulation of nonclassical states of the electromagnetic field have gained much importance in active research in quantum optics and quantum information theory \cite{pieter,braunstein05}, in the context of physical realization of quantum tasks, protocols and communications \cite{zeilinger98,bouwmeester98,bouwmeester00,nielsen00} using continuous quantum variables.
Various methods of manipulation at the single-photon level have been suggested for the preparation of nonclassical states of the optical field, based on operations of photon addition ($\hat{a}^\dag$) \cite{zavatta04} and subtraction ($\hat{a}$) \cite{wenger04} on a classical field. Agarwal and Tara \cite{agarwal91} first proposed an $m$-photon-added scheme
to create a nonclassical state from any classical state.
Zavatta \textit{et al} \cite{zavatta05} demonstrated a single photon-added coherent state by homodyne tomography technology.
A remarkable development has been the experimental realization of a general scheme by Zavatta \textit{et al} \cite{zavatta09}, based on single-photon interference, for implementing superpositions of distinct quantum operations. 
Hu \textit{et al} \cite{hu10} have recently investigated the nonclassical properties of the field states generated by subtracting any number of photons from a squeezed thermal state. Lee and Nha \cite{lee10} have studied the action of an elementary coherent superposition of $\hat{a}$ and $\hat{a}^\dag$ on continuous variable systems.

We wish to consider a general superposition of the two product (SUP) operations, $s\hat{a}\hat{a}^\dag + t\hat{a}^\dag \hat{a}$ on a classical state, where $s$ and $t$ are scalars with $s=\sqrt{1-t^2}$. This operation can be realized experimentally under suitable modification of the interference set-up proposed by Kim \textit{et al} \cite{kim08_2}. The basic unit for photon addition is a twin-photon source based on the nonlinear optical process of parametric down-conversion \cite{PDC}. The nonclassicality in the SUP operated states can be quantified and analyzed using quasiprobability distributions in the phase space. The scalars $s$ and $t$ act as control parameters for manipulation of the nonclassical character of the output state.

We observe that the SUP operation introduces nonclassicality in the classical \textit{coherent} state and there is a finite negativity in the Wigner distribution which is analyzed for different scalar parameters. The nonclassical features of the SUP operated coherent state can be further analyzed using observable features such as squeezing and sub-Poissonian statistics. In case of input \textit{thermal} state, no negativity of the Wigner function is observed. The SUP operation introduces non-Gaussianity in the classical thermal state. The nonclassical characteristics of the input thermal state is highlighted by its sub-Poissonian distribution and squeezing effect. 

This paper is organized as follows. We begin by outlining a practical SUP operation scheme in section \ref{sec2}. Various nonclassicality indicators used in our study are defined in section \ref{sec3}. In section \ref{sec4}, we present the results on SUP operated coherent and thermal states, in terms of the Wigner distribution function, a general operator-ordering parametrized quasiprobability function, Mandel's $Q$ parameter and also the quadrature squeezing parameter. The last section contains a summary of our main results.


\section{SUP operation scheme}
\label{sec2}

The generation of the desired quantum operation, $s\hat{a}\hat{a}^\dag + t\hat{a}^\dag \hat{a}$, where $\hat{a}$ and $\hat{a}^\dag$ are the annihilation and the creation operators, respectively, and $s$ and $t$ are scalars with $s=\sqrt{1-t^2}$, involves proper sequencing of photon subtraction ($\hat{a}$) and photon addition ($\hat{a}^\dag$) operators \cite{zavatta09,kim08_2}, and then coherent superposition of the ordered products by removing which-path information between them.

\begin{figure}[htb]
\begin{center}
\epsfig{figure=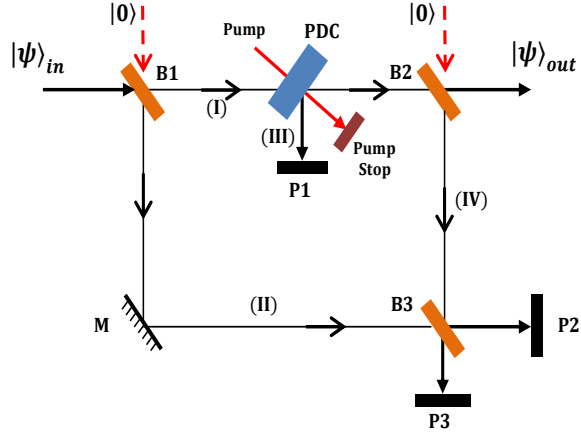, height=.24\textheight, width=.48\textwidth}
\end{center}
\caption{Experimental proposal for the generation of $s\hat{a}\hat{a}^\dag+ t\hat{a}^\dag \hat{a}$}
\label{fig1}
\end{figure}
The schematic of the experimental proposal for the generation of the quantum operation $s(\hat{a}\hat{a}^\dag)+t(\hat{a}^\dag \hat{a})$ is shown in figure~\ref{fig1}. High transmissivity beam-splitters B1 and B2 are used for photon subtraction. When an input field is incident upon a high transmissivity beam-splitter, with the other input in a vacuum mode, detection of a photon in the photo-detector implies that a photon has been subtracted ($\hat{a}$) from the incident state. The parametric down-converter (PDC) is used to add photons. A PDC produces twin photons into two different modes. If an incident field is passed through a PDC, with the other input in a vacuum state, the detection of a photon in the detector would imply that a photon (the undetected twin) has been added ($\hat{a}^\dag$) to the incident field. The operation of this optical scheme is dependent on the photo-detectors, P1, P2 and P3, which detect the success of the addition or subtraction process in an optical path. M is a highly reflective mirror. The variable transmissivity beam-splitter B3 is used to generate the desired superposition of the product states by removing the which-path information.

As shown in the schematic in figure~\ref{fig1}, the input state $|\psi\rangle_{\mathrm{in}}$ is incident upon B1. The subtraction of a photon at B1 will lead to a photon-subtracted state along path I and a photon along path II after reflection from M. In the absence of B3, the photon can be detected at P2. No detection at P2 would ensure that no subtraction has taken place at B1. The input field then proceeds to the PDC. The detection of a photon at P1, along path III, confirms the addition of a photon due to parametric down-conversion. Hence the simultaneous detection of a photon at P1 and P2 (in the absence of B3) ensures the operation $\hat{a}\hat{a}^\dag$. The photon-added field then proceeds to B2. The subtraction of a photon at B2 leads to a photon subtracted state along path I and a photon along path IV, which would be detected by P3 in the absence of B3. Hence detection at P1 and P3 (in the absence of B3), with no detection at P2 \cite{note}, ensures the operation $\hat{a}^\dag \hat{a}$. No detection at P3 would ensure that no subtraction has taken place at B2. The final output state can either be $\hat{a}\hat{a}^\dag$ or $ \hat{a}^\dag \hat{a}$ depending on the detection at P2 or P3 respectively. Hence, the two paths of detection II and IV are the two product operation indicators. The beam-splitter B3 removes this path information, and produces a superposition of the two paths and hence of the two operations.

The generation of the superposed product states can be shown mathematically using standard operators for the various paths involved in the scheme. If $|\psi\rangle_{\mathrm{in}}$ (mode I) is incident upon a high transmissivity beam-splitter B1 (transmissivity, $t_1$ $\simeq$ 1), with the other input in vacuum mode (mode II), we obtain
\begin{equation}
\hat{B}_{B1}|\psi\rangle_{\mathrm{in,I}}|0\rangle_{\mathrm{II}} \simeq (1-\frac{r_1^*}{t_1}\hat{a}_{\mathrm{I}} \hat{a}^\dag_{\mathrm{II}})|\psi_{\mathrm{in}}, 0\rangle_{\mathrm{I,II}} .
\end{equation}
The state is now incident upon a PDC with a small coupling constant $g$ and the other input in vacuum mode (III). The resulting operation can be written as
\begin{eqnarray}
&& (1-g \hat{a}_{\mathrm{I}} \hat{a}^\dag_{\mathrm{III}})\hat{B}_{B1}|\psi_{\mathrm{in}}, 0, 0\rangle_{\mathrm{I,II,III}} \nonumber \\ && \simeq (1 - g \hat{a}_{\mathrm{I}} \hat{a}^\dag_{\mathrm{III}} - \frac{r_1^*}{t_1} \hat{a}_{\mathrm{I}} \hat{a}^\dag_\mathrm{{II}} + g\frac{r_1^*}{t_1} \hat{a}_{\mathrm{I}} \hat{a}^\dag_{\mathrm{I}} \hat{a}^\dag_{\mathrm{II}} \hat{a}^\dag_{\mathrm{III})}|\psi_{\mathrm{in}},0,0\rangle_{\mathrm{I,II,III}} .
\end{eqnarray}
The photon addition occurs only when a photon is created in mode III at the PDC. Hence the state producing one photon, detected by P1, corresponds to
\[
(-g \hat{a}^\dag_{\mathrm{I}} + g\frac{r^*_1}{t_1} \hat{a}^\dag_{\mathrm{I}} \hat{a}_{\mathrm{I}} \hat{a}^\dag_{\mathrm{II}})|\psi_{\mathrm{in}},0\rangle_{\mathrm{I,II}}
\]
The state is then incident on the second high transmissivity beam-splitter B2 (transmissivity, $t_2$ $\simeq$ 1), with the other input in vacuum mode (IV). The operation leads to
\begin{eqnarray}
&&(1-\frac{r_2^*}{t_2}\hat{a}_{\mathrm{I}} \hat{a}^\dag_{\mathrm{IV}})(-g \hat{a}^\dag_{\mathrm{I}} + g \frac{r^*_1}{t_1} \hat{a}^\dag_{\mathrm{I}} \hat{a}_{\mathrm{I}} \hat{a}^\dag_\mathrm{{II}})~|\psi_{\mathrm{in}},0, 0\rangle_{\mathrm{I,II,IV}} \nonumber \\
&& \simeq (-g \hat{a}^\dag_{\mathrm{I}} + g \frac{r^*_1}{t_1} \hat{a}^\dag_{\mathrm{I}} \hat{a}_{\mathrm{I}} \hat{a}^\dag_{\mathrm{II}} - g \frac{r_2^*}{t_2}\hat{a}_{\mathrm{I}} \hat{a}^\dag_{\mathrm{I}} \hat{a}^\dag_{\mathrm{IV}} - g \frac{r_1^*}{t_1}\frac{r_2^*}{t_2}\hat{a}_{\mathrm{I}} \hat{a}^\dag_{\mathrm{I}} \hat{a}_{\mathrm{I}} \hat{a}^\dag_{\mathrm{II}} \hat{a}^\dag_{\mathrm{IV}})|\psi_{\mathrm{in}},0, 0\rangle_{\mathrm{I,II,IV}}.
\end{eqnarray}
In the absence of beam-splitter B3, photon detection of mode II at P2 (along with detection at P1 and no detection at P3) leads to the state $(g\frac{r^*_1}{t_1} \hat{a}^\dag_{\mathrm{I}} \hat{a}_{\mathrm{I}}) |\psi\rangle_{\mathrm{in}}$ and photon detection of mode IV at P3 (along with detection at P1 and no detection at P2) leads to the state $(- g \frac{r_2^*}{t_2}\hat{a}_{\mathrm{I}} \hat{a}^\dag_{\mathrm{I}} |\psi\rangle_{\mathrm{in}})$. Hence we can obtain the product state $\hat{a}^\dag_{\mathrm{I}} \hat{a}_{\mathrm{I}}$ ($\hat{a}_{\mathrm{I}} \hat{a}^\dag_{\mathrm{I}}$) using consecutive photon addition (subtraction) and subtraction (addition). Finally, we can use the beam-splitter B3 with transmissivity $t_3$ and reflectivity $r_3$ to produce the superposition state. The operation of the beam-splitter can be represented by the transformations $\acute{b} = t_3 b + r_3 c$, and $\acute{c} = t_3^*c - r_3^*b$, where $b$ and $c$ ($\acute{b}$ and $\acute{c}$) are the input (output) modes of the beam-splitter. Using the above relations, the superposition states we obtain are:
\[
(g t_3 \frac{r^*_1}{t_1} \hat{a}^\dag_{\mathrm{I}} \hat{a}_{\mathrm{I}}  - r_3 g \frac{r_2^*}{t_2} \hat{a}_{\mathrm{I}} \hat{a}^\dag_{\mathrm{I}}) |\psi\rangle_{\mathrm{in}} ,
\]
\[
(-g r_3^* \frac{r^*_1}{t_1} \hat{a}^\dag_{\mathrm{I}} \hat{a}_{\mathrm{I}}  - t_3^* g \frac{r_2^*}{t_2}\hat{a}_{\mathrm{I}} \hat{a}^\dag_{\mathrm{I}}) |\psi\rangle_{\mathrm{in}} ,
\]
which can be conveniently cast in the general form $(s\hat{a}\hat{a}^\dag+t\hat{a}^\dag \hat{a})|\psi\rangle_{\mathrm{in}}$.

\section{Nonclassicality indicators}
\label{sec3}

\noindent \textit{Wigner function}:The nonclassicality of a quantum state can be studied in terms of its phase-space distribution characterized by the Wigner distribution. For a quantum state $\hat{\rho}$, the Wigner function of the system is defined in terms of the coherent state basis \cite{scully97} as
\begin{equation}
W(\beta, \beta^*) = \frac{2}{\pi^2}e^{2|\beta|^2} \int d^2\gamma{\langle-\gamma|\hat{\rho}|\gamma\rangle e^{-2(\beta^*\gamma-\beta\gamma^*)}},
\end{equation}
where $|\gamma\rangle=\exp(-|\gamma|^2/2+\gamma \hat{a}^\dag)|0\rangle$ is a coherent state. By using the relation \cite{abramowitz72}
\begin{equation}
\sum_{n=k}^\infty n_{C_k}\,y^{n-k} = (1-y)^{-k-1},
\end{equation}
the Wigner function can be expressed in series form as \cite{moyacessa93}
\begin{equation}
W(\beta, \beta^*) = \frac{2}{\pi} \sum_{k=0}^\infty (-1)^k \langle \beta,k|\hat{\rho}|\beta,k\rangle ,
\label{eq7}
\end{equation}
where $|\beta,k\rangle$ is the usual displaced number state.

The partial negative value of the Wigner function is a one-sided condition for the nonclassicality of the related state \cite{wang64}, in the sense that one cannot conclude the state is classical when the Wigner function is positive everywhere. For example, the Wigner function of the squeezed state is Gaussian and positive everywhere but it is a well-known nonclassical state. For a classical state, a necessary but not sufficient condition is the positivity of the Wigner function. Hence a state with a negative region in the phase-space distribution is essentially nonclassical.

We may consider a generalized distribution function, viz. a parametrized quasiprobability function $\mho^{(F)}(\beta)$ describing a field state $\hat{\rho}$, defined as \cite{cahill}
\begin{eqnarray}
\label{s}
\mho^{(F)}(\beta) \equiv \frac{1}{\pi}\mathrm{Tr}\{\hat{\rho}\hat{T}^{(F)}(\beta)\},
\end{eqnarray}
where the operator $\hat{T}^{(F)}(\beta)$ is given by
$\hat{T}^{(F)}(\beta) = \frac{1}{\pi}\int \exp(\beta \xi^*-\beta^*\xi)\hat{D}^{(F)}(\xi)d^2\xi$, with $\hat{D}^{(F)}(\xi) = e^{F|\xi|^2/2}\hat{D}(\xi)$ and $\hat{D}(\xi) = e^{\xi {\hat{a}}^\dag-\xi^* \hat{a}}$. The function $\mho^{(F)}(\beta)$ can be rewritten in the number-state basis as
$\mho^{(F)}(\beta) = \frac{1}{\pi}\sum_{n,m}\rho(n,m)\langle n|\hat{T}^{(F)}(\beta)|m\rangle$,
where the matrix elements of the operator $\hat{T}^{(F)}(\beta)$ are given by
\begin{eqnarray}\nonumber
\label{s1}
\langle n|\hat{T}^{(F)}(\beta)|m\rangle & = & \left(\frac{n!}{m!}\right)^{1/2}\left(\frac{2}{1-F}\right)^{m-n+1}\left(\frac{F+1}{F-1}\right)^n{\beta^*}^{m-n}\\
& & \times\exp\left(-\frac{2|\beta|^2}{1-F}\right)L_n^{m-n}\left(\frac{4|\beta|^2}{1-F^2}\right),
\end{eqnarray}
in terms of the associated Laguerre polynomials $L_n^{m-n}(x)$. The above equation gives explicitly the $F$-dependence of $\mho^{(F)}(\beta)$. For the special values of $F = 1, 0$ and $-1$, $\mho^{(F)}(\beta)$ becomes the Glauber-Sudarshan $P$ \cite{GS}, the Wigner $W$ and the Husimi $Q$ \cite{Husimi} functions, respectively. The negativity of $\mho^{(F)}(\beta)$ for any value of the parameter $F$ indicates nonclassical nature of the state.

The nonclassical nature of a positive Wigner function can be determined using other features of the state, such as sub-Poissonian statistics and quadrature squeezing. These features, discussed later, can be attributed to the negative values which arise due to the dispersion of normally-ordered observables but are not captured by the Wigner function \cite{semenov}. In such cases, the nonclassicality is often manifested by the negativity of the $F$-parametrized distribution.

\noindent \textit{Negative Volume}: A good indicator of nonclassicality of quantum states was defined by Kenfack \textit{et al} \cite{kenfack04}. It measures the volume of the integrated negative part of the Wigner function as
\begin{equation}
V = \int\int d^2\beta |W(\beta, \beta^*)|-1.
\label{nv}
\end{equation}
By definition, this quantity $V$ is equal to zero only when the state under consideration has non-negative Wigner function. For a classical system, the Wigner distribution is positive, and the integration $\int\int d^2\beta |W(\beta, \beta^*)|$=1. For a negative Wigner function of a quantum state, the absolute value of the Wigner function can be calculated, and the above integration can be evaluated numerically.

\noindent \textit{Sub-Poissonian statistics: Mandel's Q parameter}:
The quantum character of a field can be demonstrated either in measurements of time intervals $\tau$ between detected photons demonstrating antibunching, or in photon counting measurements yielding sub-Poissonian statistics. The condition for sub-Poissonian photon statistics is given by $\langle (\Delta \hat{n})^{2} \rangle - \langle \hat{n} \rangle < 0$, which makes the normalized second-order intensity correlation function, $\gamma (0) < 1$. The states with sub-Poissonian statistics have no classical description. 

To determine the photon statistics of a single-mode radiation field, we consider Mandel's $Q$ parameter defined by \cite{mandel79}
\begin{equation}
Q \equiv \frac{\langle \hat{a}^{\dag 2} \hat{a}^2 \rangle - \langle \hat{a}^\dag \hat{a}\rangle^2}{\langle \hat{a}^\dag \hat{a} \rangle} .
\end{equation}
$Q=0$ stands for Poissonian photon statistics. $Q<0$ corresponds to the case of sub-Poissonian distribution. This means that a nonclassical state often shows negative $Q$ values.

\noindent \textit{Squeezing}: The quadrature squeezing of a field can be used to study its nonclassical properties. To analyze the squeezing properties of the radiation field, we introduce two hermitian quadrature operators
\begin{equation}
\hat{X}=\hat{a}+\hat{a}^\dag,~~~~~~\hat{Y}=-i(\hat{a}-\hat{a}^\dag).
\end{equation}
These two quadrature operators satisfy the commutation relation $[\hat{X}, \hat{Y}]=2i$, and, as a result, the uncertainty relation $(\Delta \hat{X})^2(\Delta \hat{Y})^2\geq 1$. A state is said to be squeezed if either $(\Delta \hat{X})^2$ or $(\Delta \hat{Y})^2$ is less than its coherent state value. To review the principle quadrature squeezing \cite{luks88}, we define an appropriate quadrature operator \cite{wang03}
\begin{eqnarray}
\hat{X}_\theta = \hat{X}\cos\theta+\hat{Y}\sin\theta = \hat{a}e^{-i\theta}+\hat{a}^\dag e^{i\theta}.
\end{eqnarray}
The squeezing of $\hat{X}_\theta$ is characterized by the condition $\langle:(\Delta \hat{X}_\theta)^2:\rangle<0$, where the double dots denote the normal ordering of operators. After expanding the terms in $\langle:(\Delta \hat{X}_\theta)^2:\rangle$ and minimizing its value over the whole angle $\theta$, one gets \cite{lee10}
\begin{eqnarray}
S_{\mathrm{opt}} & = & \langle:(\Delta \hat{X}_\theta)^2:\rangle_{\mathrm{min}} \nonumber \\
& = & -2|\langle \hat{a}^{\dag 2}\rangle-\langle \hat{a}^\dag\rangle^2|+2\langle \hat{a}^\dag \hat{a}\rangle-2|\langle \hat{a}^\dag \rangle|^2 .
\label{eq14}
\end{eqnarray}
The nonclassical states correspond to the negative values of $S_{\mathrm{opt}}$, $-1\leq S_{\mathrm{opt}}<0$.

The negativity of the $Q$ function and squeezing are not necessary conditions for identifying nonclassical regimes of quantum states but are sufficient ones. In different regimes of the scalars $s$ and $t$, the nonclassicality of the SUP operated states are exhibited via these indicators. As mentioned earlier, the $Q$ function and the squeezing parameter $S$ can be negative and hence nonclassical for states that have a positive Wigner function, as shown in \cite{semenov, cessa, janszky96}. Conversely, there are also instances where states with partial negative Wigner functions have positive $Q$ functions \cite{OC}.

\section{Results}
\label{sec4}
\subsection{SUP operated coherent state}

Let the density matrix of an arbitrary quantum input state of the single-mode radiation field be
\begin{equation}
\hat{\rho}_{\mathrm{in}} = \sum_{m=0}^\infty \sum_{n=0}^\infty \rho(m, n)|m\rangle\langle n|.
\end{equation}
The output state, produced by applying the SUP operator ($s\hat{a}\hat{a}^\dag + t\hat{a}^\dag \hat{a}$) on $\hat{\rho}_{\mathrm{in}}$, is given by
\begin{equation}
\hat{\rho}_{\mathrm{out}} = \frac{1}{N}[s(\hat{a}\hat{a}^\dag)+t(\hat{a}^\dag \hat{a})]\hat{\rho}_{\mathrm{in}}[s(\hat{a}\hat{a}^\dag)+t(\hat{a}^\dag \hat{a})],
\label{eq2}
\end{equation}
where $N$ is the normalization constant.

If we consider our input single-mode radiation field to be in a coherent state,
\begin{equation}
|\alpha\rangle =\exp\left(\frac{-|\alpha|^2}{2}\right)\sum_{n=0}^{\infty} \frac{\alpha^{n}}{\sqrt{n!}}|n\rangle ,
\end{equation}
where $|\alpha|^2$ is the average photon number, the SUP operated coherent state (SOCS) is given by
\begin{equation}
\hat{\rho}_{\mathrm{coh}} = N_1^{-1}[s(\hat{a}\hat{a}^\dag)+t(\hat{a}^\dag \hat{a})]|\alpha\rangle\langle\alpha|[s(\hat{a}\hat{a}^\dag)+t(\hat{a}^\dag \hat{a})] ,
\label{eq3}
\end{equation}
where $N_1 = s^2+(s+t)(3s+t)|\alpha|^2+(s+t)^2|\alpha|^4$ is the normalization constant.

To analyze the nonclassicality of the SOCS, we need to obtain the phase-space distribution of the density matrix in terms of the Wigner function. The expression for the Wigner function in the series form is given in (\ref{eq7}). For SOCS density matrix, the displaced number state expectation value is given by the relation
\begin{eqnarray}
& & \langle\beta,k|(s\hat{a}\hat{a}^\dag+t\hat{a}^\dag \hat{a})|\alpha\rangle \nonumber \\ & & = \langle k|D^\dag(\beta)(s\hat{a}\hat{a}^\dag+t\hat{a}^\dag
\hat{a})|\alpha\rangle \nonumber \\ & & = \langle k|\{s(\hat{a}+\beta)(\hat{a}^\dag+\beta^*)+ t(\hat{a}^\dag+\beta^*)(\hat{a}+\beta)\}D^\dag(\beta)|\alpha \rangle .
\label{eq8}
\end{eqnarray}
Substituting the above expression (\ref{eq8}) into the general expression (\ref{eq7}), the Wigner function for SOCS is obtained as
\begin{eqnarray}
W_{\mathrm{SOCS}}(\beta, \beta^*) & = & W_{\mathrm{coh}}(\beta, \beta^*) N_1^{-1}\big[{M_1}^2+2(s+t)M_1(\alpha^*\beta+\alpha\beta^*) \nonumber \\
& & +(s+t)^2|\alpha|^2(4|\beta|^2-1)\big] ,
\end{eqnarray}
where $W_{\mathrm{coh}}(\beta, \beta^*)=\frac{2}{\pi} e^{-2|\beta-\alpha|^2}$ is the Wigner function of the input coherent state $|\alpha\rangle$,\,$M_1=s-(s+t)|\alpha|^2$, $N_1 = s^2+(s+t)(3s+t)|\alpha|^2+(s+t)^2|\alpha|^4$.

\begin{figure}[ht]
\centering
\subfigure{\includegraphics[width=8cm]{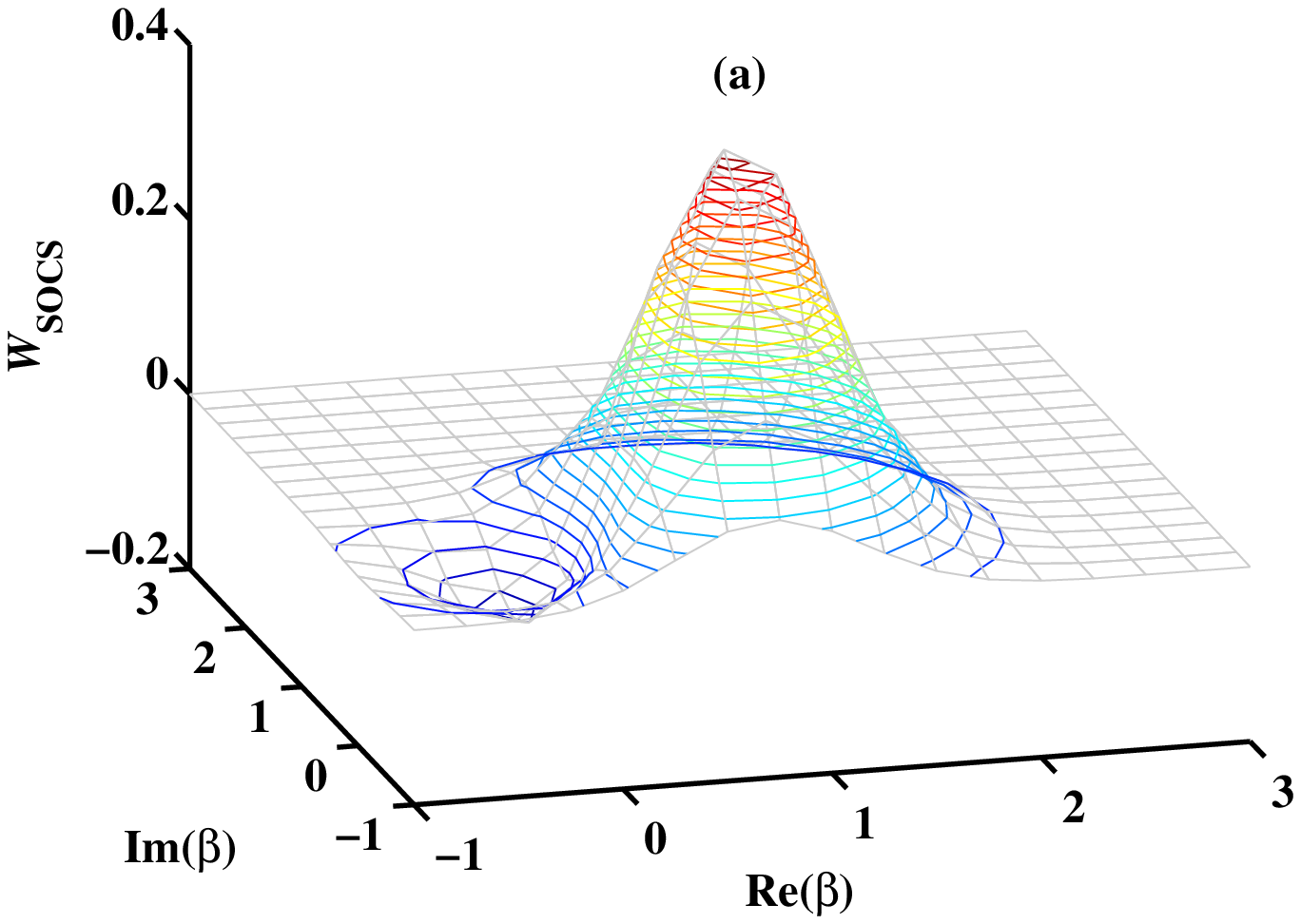}}\hspace{0.5cm}
\subfigure{\includegraphics[width=7cm]{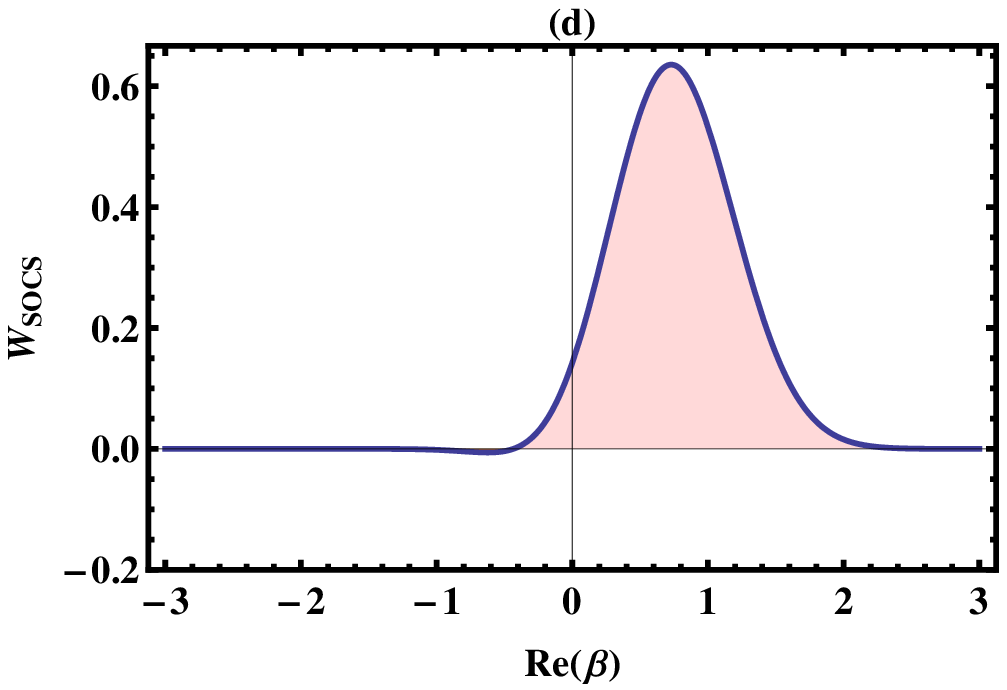}}\hspace{0.5cm}
\subfigure{\includegraphics[width=8cm]{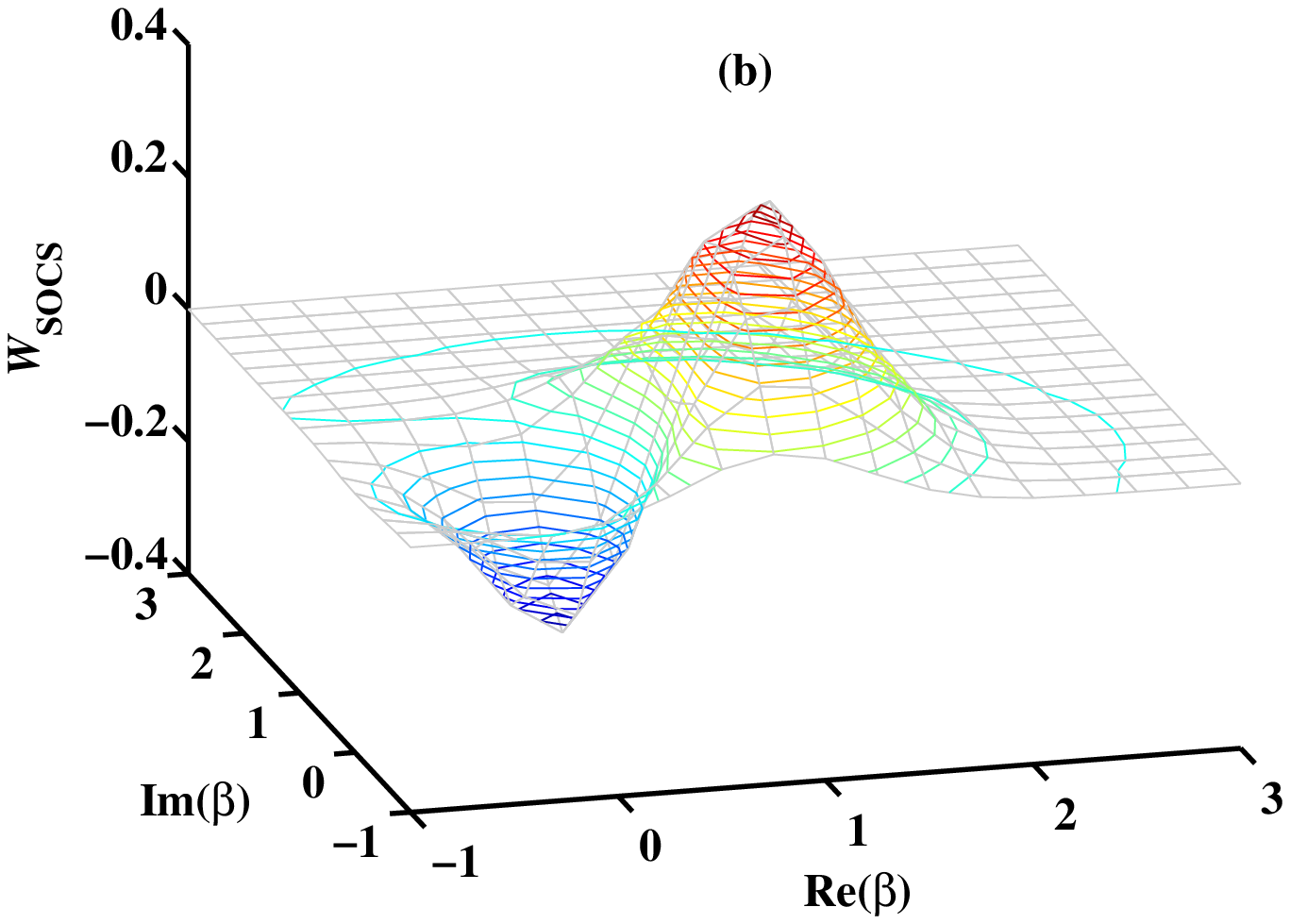}}\hspace{0.5cm}
\subfigure{\includegraphics[width=7cm]{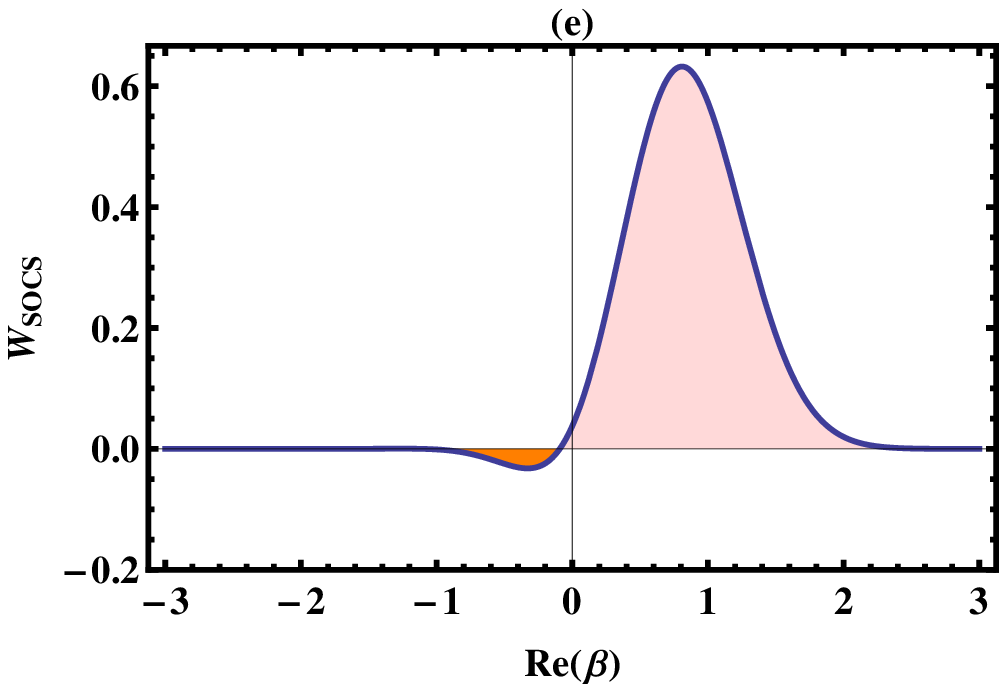}}\hspace{0.5cm}
\subfigure{\includegraphics[width=8cm]{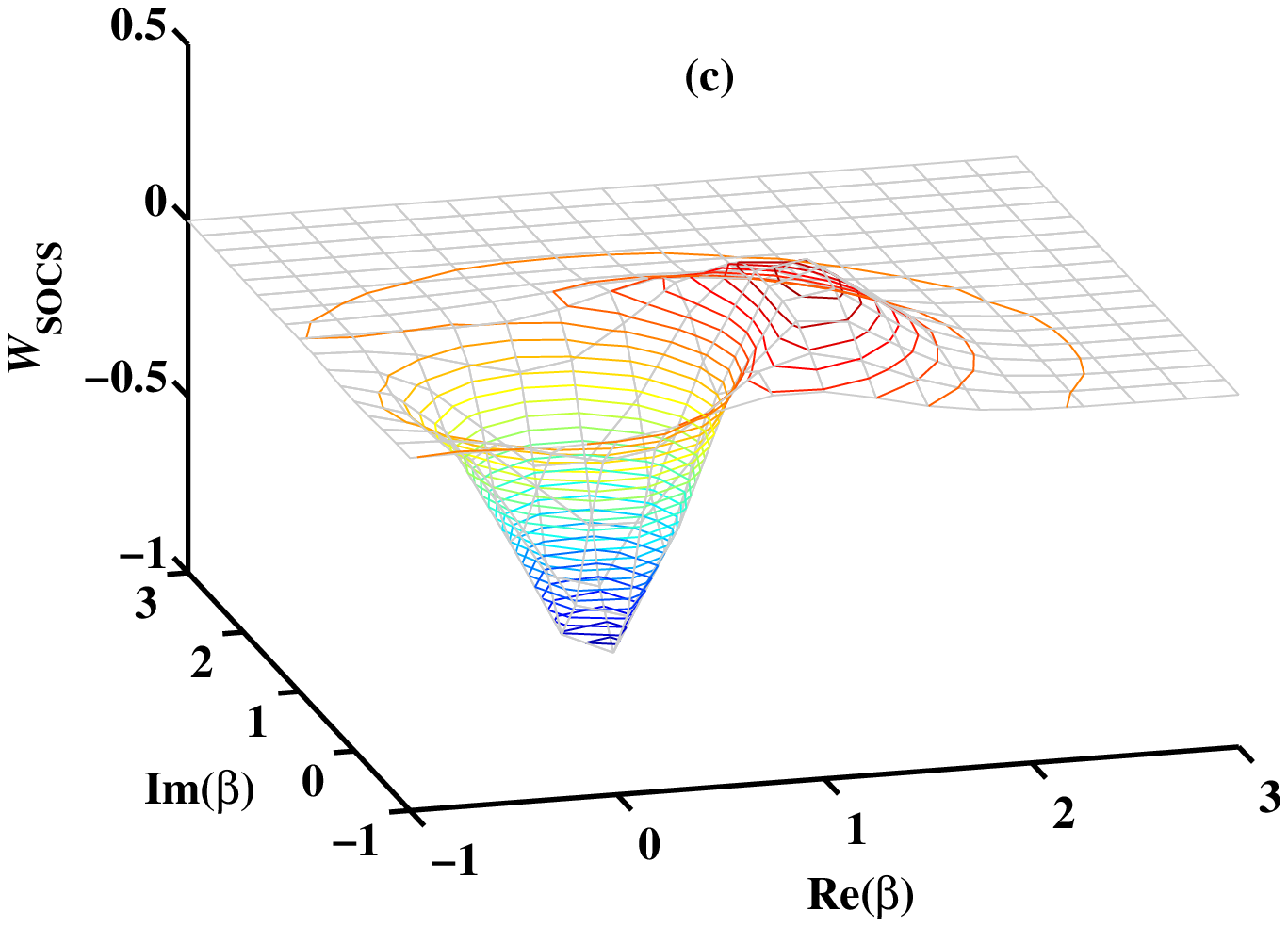}}\hspace{0.5cm}
\subfigure{\includegraphics[width=7cm]{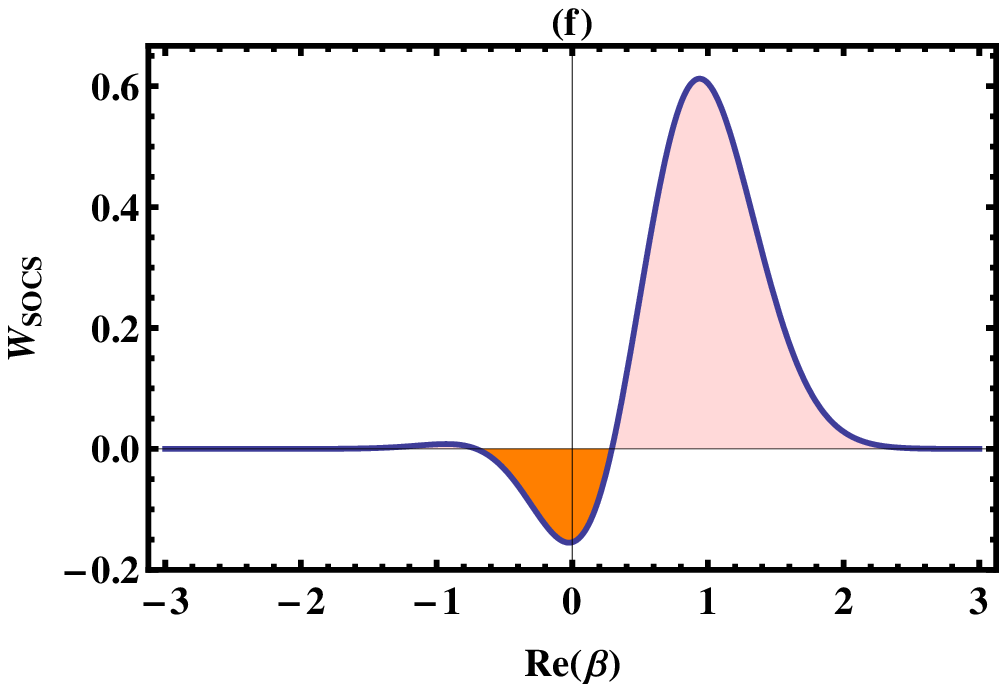}}\hspace{0.5cm}
\caption{Wigner function $W_{\mathrm{SOCS}}$ of the state $(s\hat{a}\hat{a}^\dag+t\hat{a}^\dag \hat{a})|\alpha\rangle$, where $s=\sqrt{1-t^2}$, on the left panel as a function of Re$(\beta)$ and Im$(\beta)$ for $|\alpha| =0.4$: (a) $t=0.1$, (b) $t=0.5$, and (c) $t=0.9$. (d), (e) and (f) on the right are the 2D plots corresponding to the Wigner functions (a), (b) and (c), respectively, as a function of Re$(\beta)$ with Im$(\beta) = 0$. The negativity in the phase-space distribution is clearly seen.}
\label{fig2}
\end{figure}

Figure \ref{fig2} represents the Wigner distribution, $W_{\mathrm{SOCS}}$, in the phase space for fixed value of $|\alpha|$ and different values of the scalar parameters $s$ and $t$. The Wigner distribution is plotted as a function of Re$(\beta)$ and Im$(\beta)$ in the phase space. The plots on the right of figure~\ref{fig2} are the 2D plots of the Wigner distribution varying with Re$(\beta)$ (Im$(\beta)$=0). The negative dip of the Wigner distribution increases with $t$. It is clear that performing the SUP operation transforms a purely classical coherent state to a nonclassical one in terms of the negativity of the Wigner distribution \cite{comment}.

Let us analyze the area of negativity of the Wigner function, $W_{\mathrm{SOCS}}(\beta, \beta^*)$, for selected scalar parameters. For $t=1$, the Wigner distribution is given by the relation
\begin{eqnarray}
W_{\mathrm{SOCS}}(\beta, \beta^*) & = & \frac{8}{\pi}(a^2+b^2) e^{-2\{(x-a)^2+(y-b)^2\}} \nonumber \\
& & \times \left[\left(x-\frac{a}{2}\right)^2+\left(y-\frac{b} {2}\right)^2-\frac{1}{4} \right] ,
\end{eqnarray}
where $\beta=x+iy$, $\alpha=a+ib$. The negative region is represented by
$\left\{\left(x-\frac{a}{2}\right)^2+\left(y-\frac{b}{2}\right)^2-\frac{1}{4}\right\}<0$, bounded by a circle of radius 1/2 and centered at $\left( \frac{a}{2}, \frac{b}{2} \right)$.

For $t=1/\sqrt{2}$, the Wigner distribution is
\begin{eqnarray}
W_{\mathrm{SOCS}}(\beta, \beta^*) & = &
W_{\mathrm{coh}}(\beta, \beta^*) s^2\big[1-8|\alpha|^2+16|\alpha|^2|\beta|^2+4|\alpha|^4 \nonumber \\
& & +4(1-2|\alpha|^2)(\alpha^*\beta+\alpha\beta^*)\big].
\end{eqnarray}
Thus the area of negativity is bounded by a circular region
$\left\{x-\left(\frac{a}{2}-\frac{a}{4(a^2+b^2)}\right)\right\}^2+\left\{y-\left(\frac{b}{2}
-\frac{b}{4(a^2+b^2)}\right)\right\}^2 = \frac{1}{4}$.

For $t=0$, the Wigner function is
\begin{eqnarray}
W_{\mathrm{SOCS}}(\beta, \beta^*) & = & W_{\mathrm{coh}}(\beta, \beta^*) s^2\big[1-3|\alpha|^2+4|\alpha|^2|\beta|^2+|\alpha|^4 \nonumber \\
& & +2(1-|\alpha|^2)(\alpha^*\beta+\alpha\beta^*)\big] ,
\end{eqnarray}
and the corresponding negative region is again within a circle,
$\left\{x-\left(\frac{a}{2}-\frac{a}{2(a^2+b^2)}\right)\right\}^2+\left\{y-\left(\frac{b}{2} -\frac{b}{2(a^2+b^2)}\right)\right\}^2 = \frac{1}{4}$.

Therefore, in all the cases, the negative region becomes a circle of radius $1/2$, i.e. the negative area is independent of the choice of the scalar $t$.
The nonclassical nature of the SOCS is evident from the negative region of the Wigner function; however, the degree of quantumness cannot be quantified for different scalar parameters.

The nonclassicality of SOCS can also be investigated using the $F$-parametrized quasiprobability function. For the SOCS density matrix (\ref{eq3}), using the relation for $\mho^{(F)}$ (\ref{s}), we obtain
\begin{eqnarray}\nonumber
& &\mho^{(F)}_{\mathrm{coh}}(\beta) =  \frac{1}{\pi}N_1^{-1}\left[s^2 \langle\alpha|\hat{T}^{(F)}(\beta)|\alpha\rangle+s(s+t)\alpha^* e^{-|\alpha|^2}\sum_n\sqrt{n+1}\frac{(|\alpha|^2)^n}{n!}\right.\\\nonumber
& & \left.\langle n+1|\hat{T}^{(F)}(\beta)|n\rangle+s(s+t)\alpha e^{-|\alpha|^2}\sum_n\sqrt{n+1}\frac{(|\alpha|^2)^n}{n!}\langle n|\hat{T}^{(F)}(\beta)|n+1\rangle\right.\\\nonumber
& & \left.+(s+t)^2|\alpha|^2 e^{-|\alpha|^2}\sum_n(n+1)\frac{(|\alpha|^2)^n}{n!}\langle n+1|\hat{T}^{(F)}(\beta)
|n+1\rangle\right],
\end{eqnarray}
where $\langle\alpha|\hat{T}^{(F)}(\beta)|\alpha\rangle = \left(\frac{2}{1-F}\right)\exp\left[-\left(\frac{2}{1-F}\right)|\beta-\alpha|^2\right]$
and $\langle n|\hat{T}^{(F)}(\beta)|m\rangle$ is given by (\ref{s1}).

\begin{figure}[ht]
\centering
\includegraphics{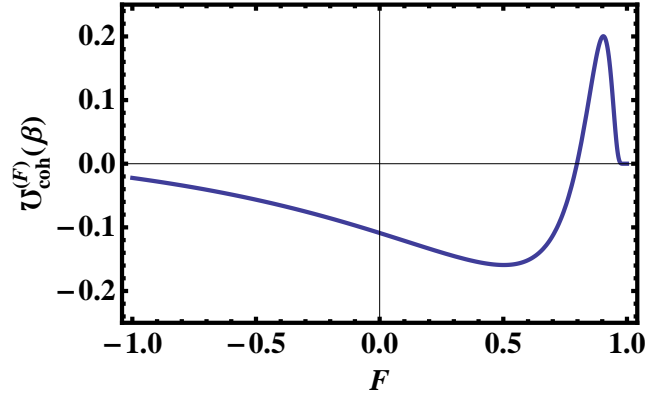}
\caption{$F$-parametrized quasiprobability function $\mho^{(F)}_{\mathrm{coh}}(\beta)$ of a coherent state $|\alpha\rangle$ after the operation $(s\hat{a}\hat{a}^\dag+t\hat{a}^\dag \hat{a})$, where $s=\sqrt{1-t^2}$, as a function of $F$ for $|\beta| = 0.8$, $|\alpha|= 0.4$ and  $t = 0.5$.}
\label{pars}
\end{figure}
In figure~(\ref{pars}) we observe the behavior of the $F$-parametrized function for different values of the parameter $F$. The function is negative around $F$=0, which is consistent with our observed nonclassicality of the Wigner function.\\

The nonclassical nature of SOCS can also be captured by its negative volume (\ref{nv}).
\begin{figure}[ht]
\centering
\includegraphics[width=8cm]{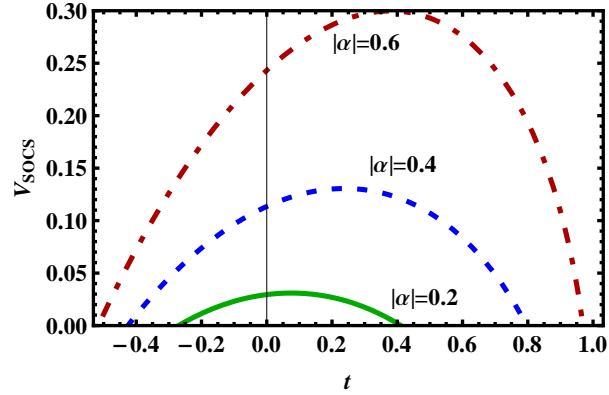}
\caption{Negative volume as a function of $t$ for a coherent state input with $|\alpha| =$ 0.2 (continuous), 0.4 (dashed) and 0.6 (dot-dashed).}
\label{fig3}
\end{figure}
Figure \ref{fig3} clearly shows that, for fixed $t$, the negative volume $V_{\mathrm{SOCS}}$ increases with the amplitude $|\alpha|$ of the input coherent field.
$V_{\mathrm{SOCS}}$ first increases and then decreases with increasing $t$. The range of $t$ for which SOCS shows nonclassicality is dependent on the value of $|\alpha|$.

The sub-Poissonian statistics of SOCS can be established by using the following:
\begin{equation}
Q_{\mathrm{SOCS}} = -\frac{|\alpha|^2}{K_1}N_1^{-1}\left\{K_1(K_1-1)-(s+t)^2 (2|\alpha|^2+3) - 2s(s+t) \right\} ,
\end{equation}
where $K_1=\left[(s+t)|\alpha|^2+(2s+t)\right]^2+(s+t)^2|\alpha|^2$.
\begin{figure}[ht]
\centering
\subfigure{\includegraphics[width=8cm]{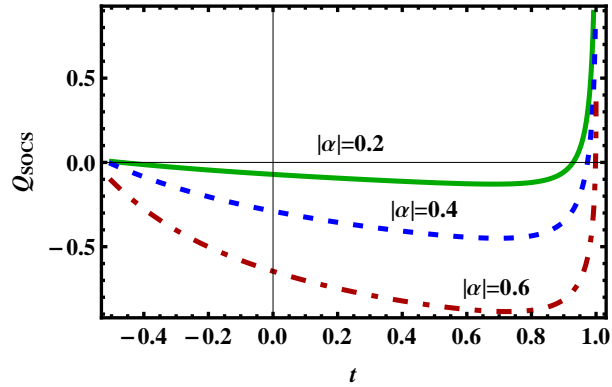}}
\caption{Mandel's $Q$ parameter as a function of $t$ and with $|\alpha| =$ 0.2 (continuous), 0.4 (dashed) and 0.6 (dot-dashed), for a coherent state input.}
\label{fig4}
\end{figure}

In order to see the variation of the $Q$ parameter with $|\alpha|$ (coherent field), we plot the $Q$ function against the parameter $t$ in figure~\ref{fig4}. $Q$ exhibits the sub-Poissonian character for the coherent input state and increases its negativity as $|\alpha|$ increases. But at $t=1$, the $Q$ parameter suddenly changes its characteristics to indicate super-Poissonian distribution.

The squeezing parameter for SOCS is calculated to yield the following:
\begin{equation}
S_{\mathrm{SOCS}} = 2N_1^{-1}(s+t)^2|\alpha|^2 .
\label{sqz}
\end{equation}
\begin{figure}[ht]
\centering
\includegraphics[width=5cm,height=8cm,angle=-90]{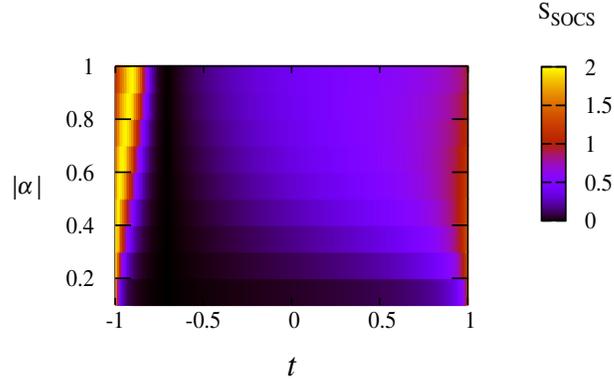}
\caption{Contour plot for $S_{\mathrm{SOCS}}$ as a function of $t$ and $|\alpha|$ for an input coherent state.}
\label{fig5}
\end{figure}
From equation (\ref{sqz}), we can see that $S_{\mathrm{SOCS}}$ is positive for all $t$ and $|\alpha|$. The superposed product (SUP) operation cannot inject squeezing property into the coherent state character (see figure~\ref{fig5}).

\begin{figure}[ht]
\centering
\includegraphics[width=8cm]{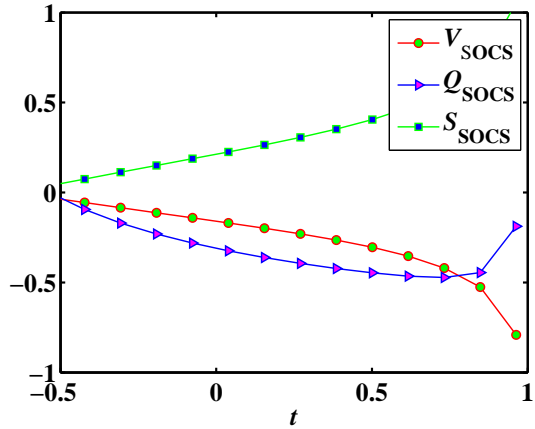}
\caption{A comparison of the different nonclassical indicators for SOCS as a function of the scalar parameter $t$ with $|\alpha|=0.4$.}
\label{fig6}
\end{figure}

In general, the nonclassicality indicators can be compared to observe the signature of quantumness introduced by the SUP operation on the input coherent field. We observe from the phase space distribution, the negative region of the Wigner distribution is an indicator of nonclassicality of the state but cannot quantify the degree of nonclassicality. Another indicator of nonclassicality, viz. negative volume of the Wigner function
is able to quantitatively classify the nonclassicality. The negative volume decreases and the $Q$ parameter increases when $t$ becomes close to 1. The squeezing parameter $S$ does not exhibit any nonclassicality and hence fails as an indicator. A comparison of the indicators is shown in figure~\ref{fig6}.

The generated nonclassical SOCS states can prove to be useful in a wide variety of tasks and applications, with optical components effecting Gaussian processes being readily available in the laboratory. The experimental and theoretical developments on continuous-variable quantum information processes in the Gaussian realm can be found in a recent review \cite{RMPloyd}.

\subsection{SUP operated thermal state}

If we consider an input single-mode thermal field with frequency $\omega$ and at absolute temperature $T$, with the Fock state representation
\begin{equation}
\hat{\rho}_{\mathrm{in}}=\frac{1}{(1+\bar{n})}\sum_n\left(\frac{\bar{n}}{1+\bar{n}}\right)^n|n\rangle\langle n| ,
\end{equation}
where $\bar{n}=[e^{\hbar \omega/kT}-1]^{-1}$ is the average photon number, $k$ being the Boltzmann constant, the resulting SUP operated thermal state (SOTS) is given by
\begin{equation}
\hat{\rho}_{\mathrm{th}} = N_2^{-1}
[s(\hat{a}\hat{a}^\dag)+t(\hat{a}^\dag \hat{a})]\hat{\rho}_{\mathrm{in}}[s(\hat{a}\hat{a}^\dag)+t(\hat{a}^\dag \hat{a})] ,
\label{therm}
\end{equation}
where $N_2 = s^2(1+\bar{n})(1+2\bar{n})+4st\bar{n}(1+\bar{n})+t^2\bar{n}(1+2\bar{n})$ is the normalization constant.

Using the series expression (\ref{eq7}) for the Wigner function, the phase space distribution for the thermal state is
\begin{equation}
W_{\mathrm{SOTS}}(\beta, \beta^*) = W_{\mathrm{th}}(\beta, \beta^*)N_2^{-1}\big[(M_2+s)^2+(s+t)M_2\big] ,
\end{equation}
where $W_{\mathrm{th}}(\beta, \beta^*)=\frac{2} {\pi}\frac{1}{(1+2\bar{n})}e^{-\frac{2|\beta|^2}{1+2\bar{n}}}$ is the Wigner function of the input thermal state, $M_2=\frac{4\bar{n}(1+\bar{n})}{(1+2\bar{n})^2}(s+t)|\beta|^2$, $N_2 = s^2(1+\bar{n})(1+2\bar{n})+4st\bar{n}(1+\bar{n})+t^2\bar{n}(1+2\bar{n})$.
\begin{figure}[ht]
\centering
\subfigure{\includegraphics[width=8cm]{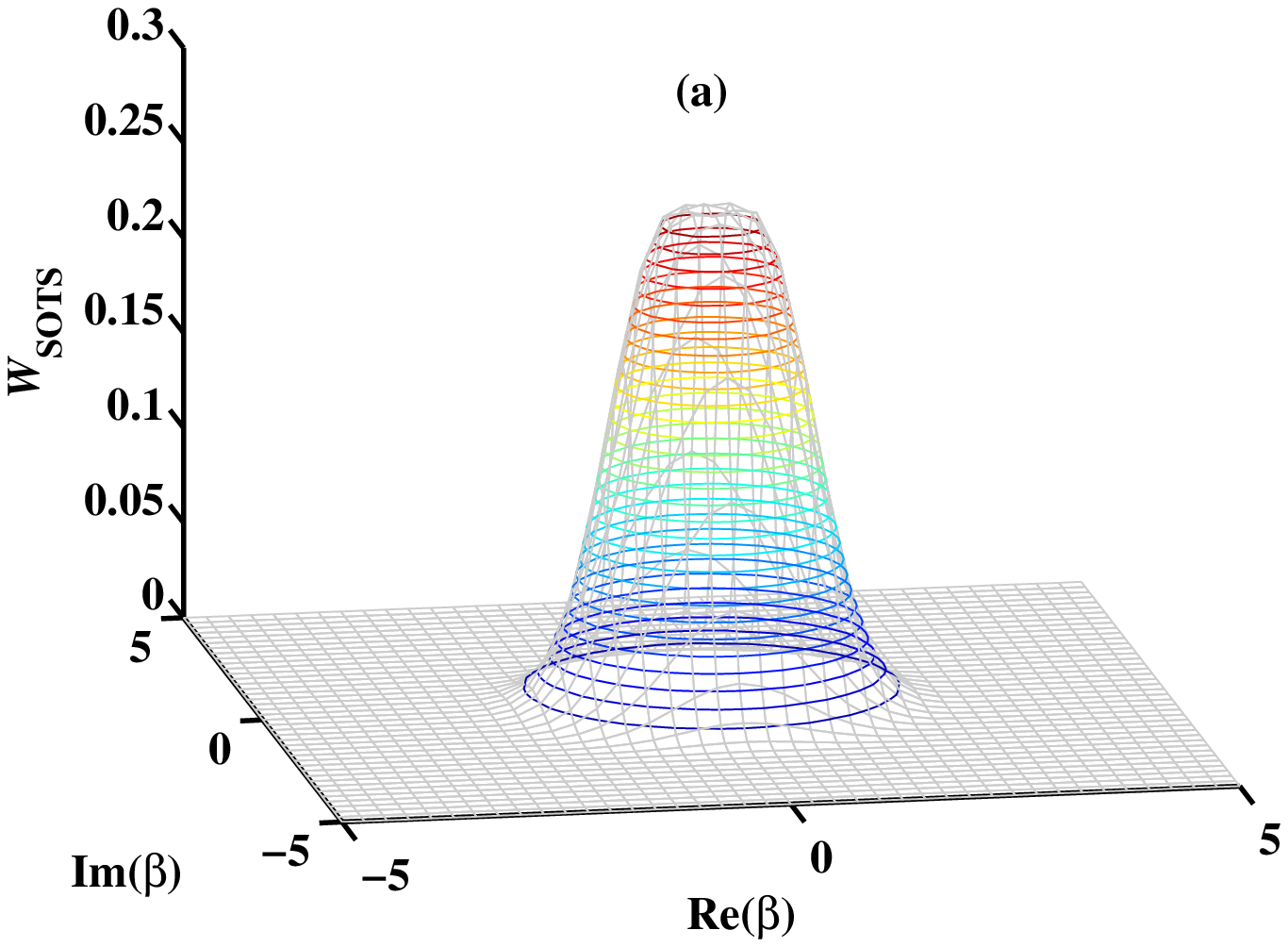}}\hspace{0.5cm}
\subfigure{\includegraphics[width=7cm]{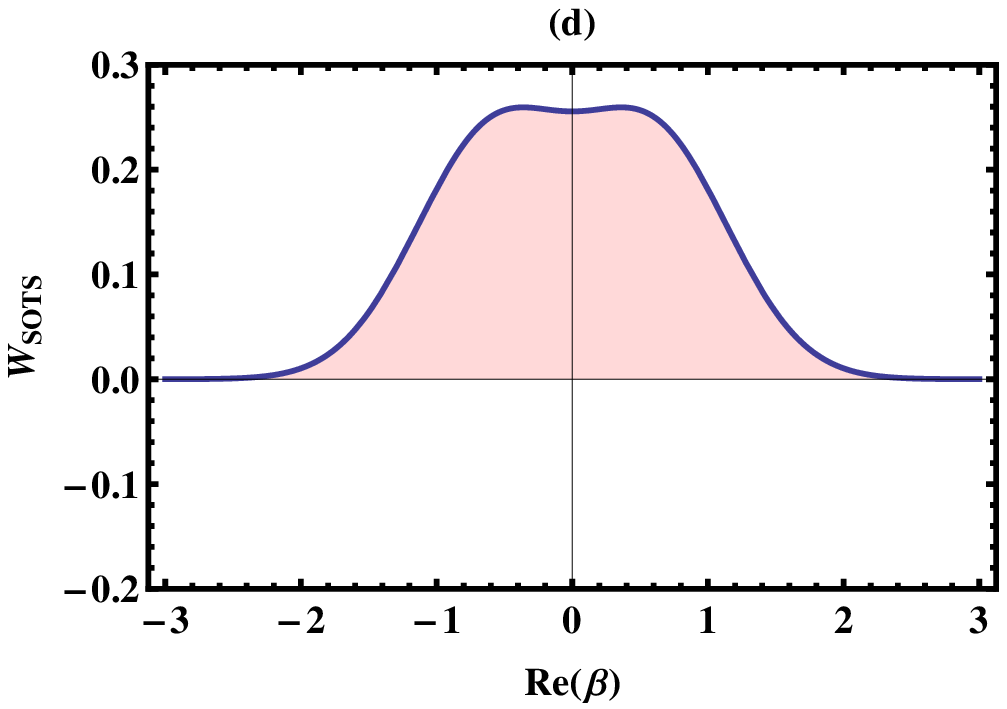}}\hspace{0.5cm}
\subfigure{\includegraphics[width=8cm]{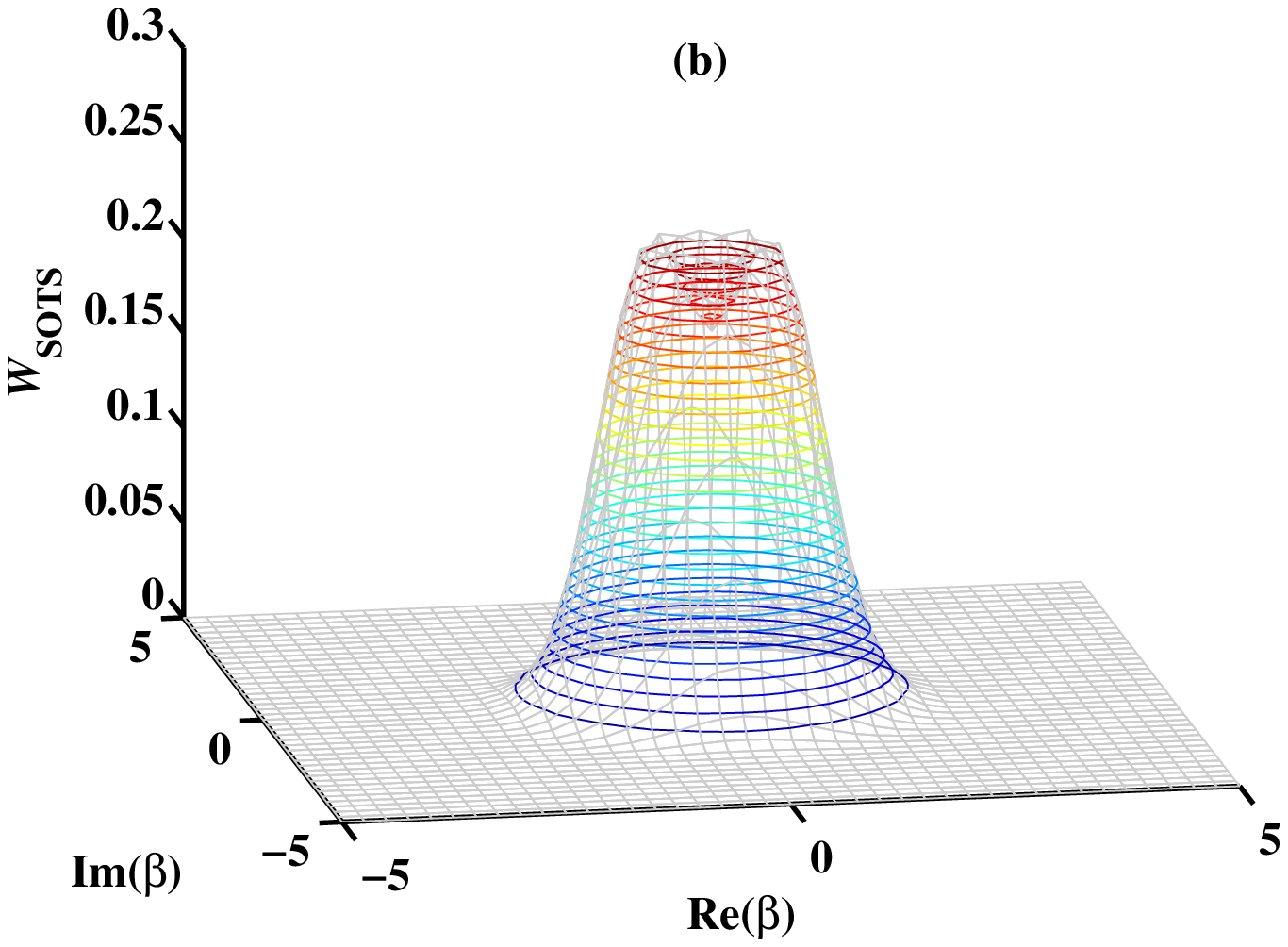}}\hspace{0.5cm}
\subfigure{\includegraphics[width=7cm]{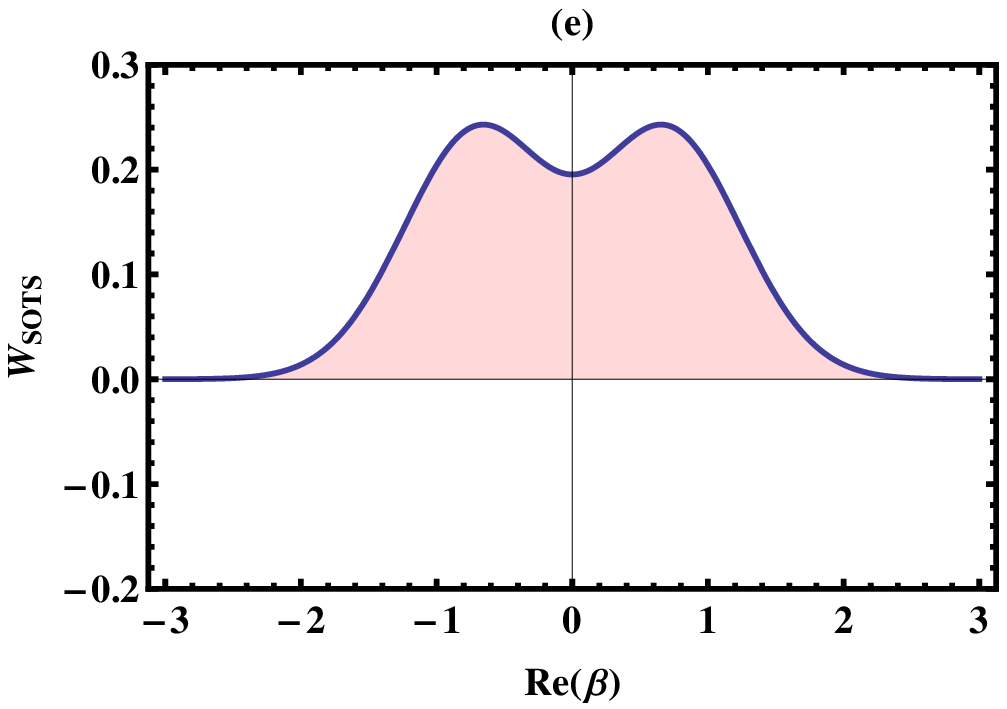}}\hspace{0.5cm}
\subfigure{\includegraphics[width=8cm]{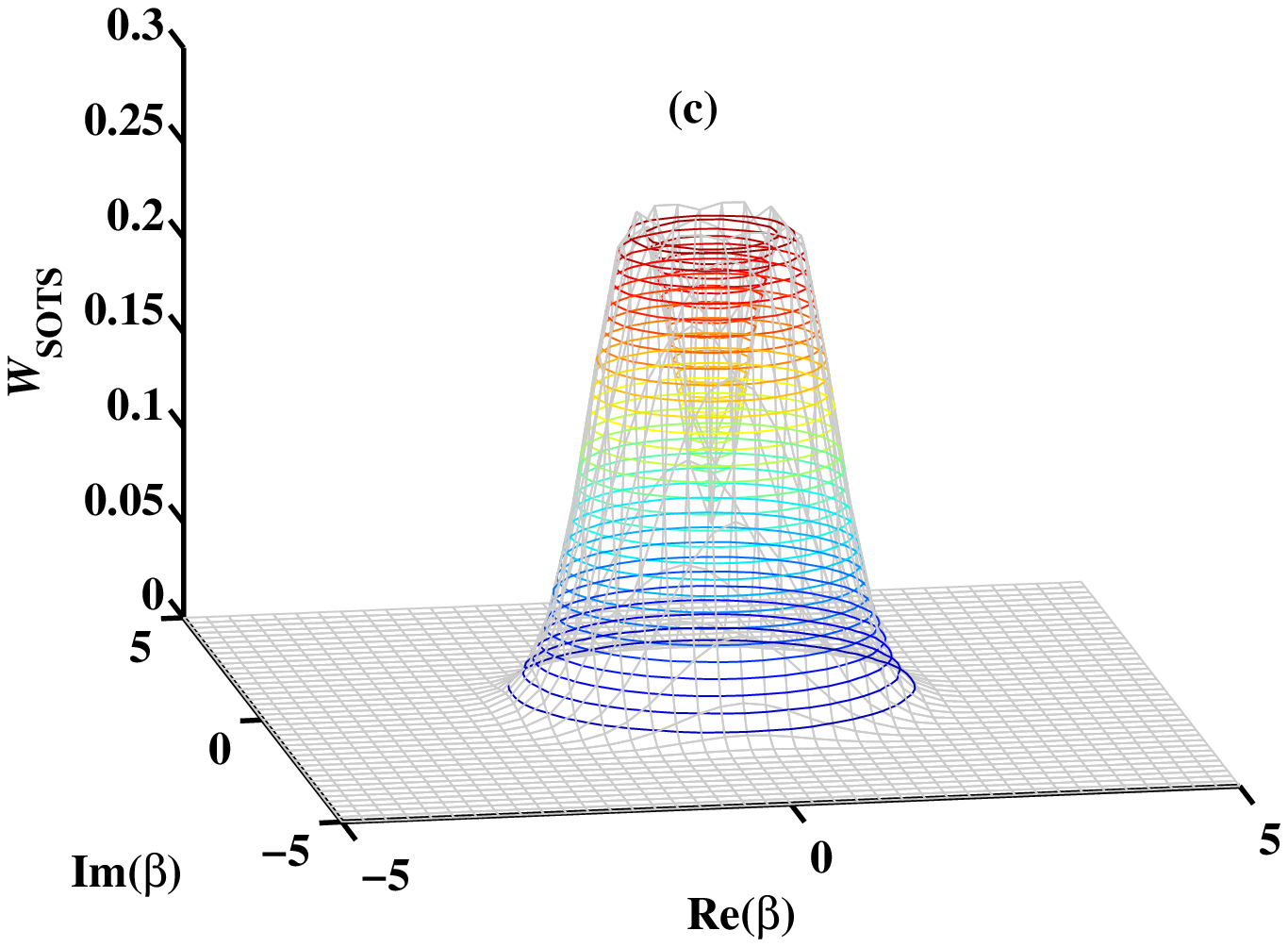}}\hspace{0.5cm}
\subfigure{\includegraphics[width=7cm]{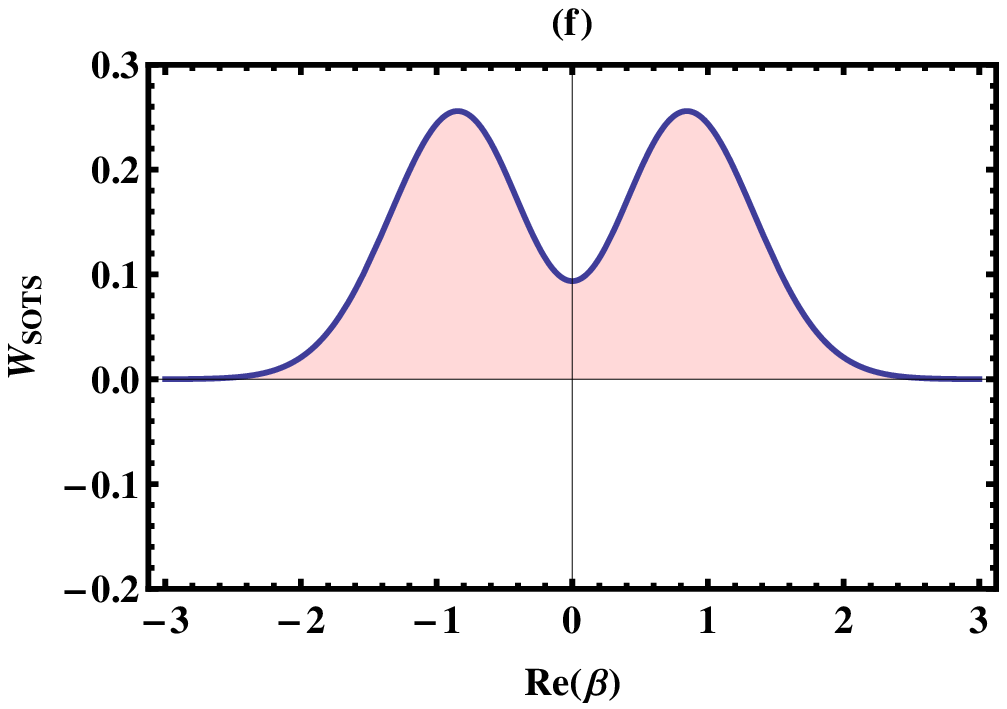}}\hspace{0.5cm}
\caption{Wigner function $W_{\mathrm{SOTS}}$ of a thermal state after the operation $s(\hat{a}\hat{a}^\dag)+t(\hat{a}^\dag \hat{a})$, where $s=\sqrt{1-t^2}$, as a function of Re$(\beta)$ and Im$(\beta)$ for $\bar{n}=0.2$: (a) $t=0.1$, (b) $t=0.5$, (c) $t=0.9$. (d), (e) and (f) on the right are the 2D plots corresponding to the Wigner functions (a), (b) and (c) respectively, as a function of Re$(\beta)$ with Im$(\beta) = 0$. There is no negativity seen in the phase-space distribution.}
\label{fig7}
\end{figure}
In figure~\ref{fig7}, we plot the Wigner distribution $W_{\mathrm{SOTS}}(\beta, \beta^*)$ as a function of $t$ for fixed $\bar{n}=0.2$. Unlike the case of input coherent state, $W_{\mathrm{SOTS}}$ has no negative region but the Wigner function does not remain Gaussian. Hence, $s(\hat{a}\hat{a}^\dag)+t(\hat{a}^\dag \hat{a})$ generates a non-Gaussian state from the Gaussian thermal state. The nonclassical nature of SOTS cannot be captured by the Wigner function, which remains positive. However this is not a necessary condition for the nonclassicality of SOTS. 

The nonclassicality of the state can be investigated using the $F$-paramterized quasiprobability function (\ref{s}).
For the SUP operated thermal state, using the expression for the output density matrix (\ref{therm}), we obtain
\begin{eqnarray}\nonumber
\mho^{(F)}_{\mathrm{th}}(\beta) & = & \frac{1}{\pi}N_2^{-1}\left(\frac{1}{1+\bar{n}}\right)\left(\frac{2}{1-F}\right)\exp\left(-\frac{2|\beta|^2}{1-F}\right)\\
& & \times\sum_n [s+(s+t)n]^2\left\{\left(\frac{\bar{n}}{1+\bar{n}}\right)\left(\frac{F+1}{F-1}\right)\right\}^n
L_n\left(\frac{4|\beta|^2}{1-F^2}\right) .
\end{eqnarray}


\begin{figure}[ht]
\centering
\includegraphics[width=7.9cm]{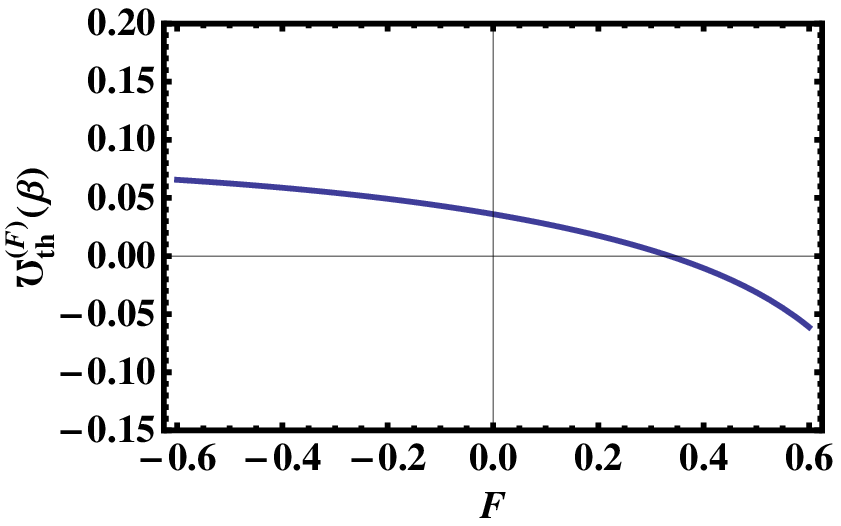}
\caption{$F$-parametrized quasiprobability function $\mho^{(F)}_{\mathrm{th}}(\beta)$ of a thermal state after the operation $(s\hat{a}\hat{a}^\dag+t\hat{a}^\dag \hat{a})$, where $s=\sqrt{1-t^2}$, as a function of $F$ for $|\beta| = 0.5$, $\bar{n} = 0.2$ and $t = 0.9$.}
\label{fig7.1}
\end{figure}

In figure~\ref{fig7.1}, we plot $\mho^{(F)}_{\mathrm{th}}(\beta)$ as a function of $F$. For $t = 0.9$, the $F$-parametrized quasiprobability is positive for $F < 0$ and becomes negative for $F > 0.3$. The function is always positive at $F = 0$ matching with non-negative Wigner function for SOTS. $\mho^{(F)}_{\mathrm{th}}(\beta)$ becomes highly negative as $F \rightarrow 1$. Hence, the nonclassicality of SOTS can be evidenced by the negativity of the normally ordered Glauber-Sudarshan $P$ function. Similar behavior is also observed for other values of the parameter $t$.

To further check the nonclassical effects of SUP operations on input thermal state, we study the sub-Poissonian statistics of SOTS. Mandel's $Q$ parameter for SOTS is found to be
\begin{equation}
Q_{\mathrm{SOTS}} = -\frac{\bar{n}}{K_2}N_2^{-1}\left\{ K_2(K_2-3)-6(\bar{n}+1)^2(s+t)^2+t^2 \right\} ,
\end{equation}
where $K_2=2(\bar{n}+1)(s+t)\left[3\bar{n}(s+t)+2s\right]+t^2$.
\begin{figure}[ht]
\centering
\includegraphics[width=8cm]{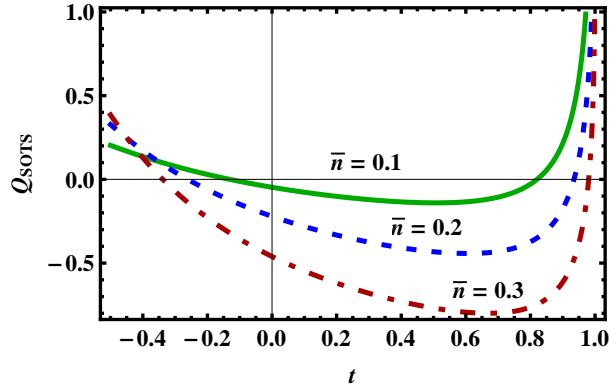}
\caption{Mandel's $Q$ parameter as a function of $t$ and with $\bar{n} =$ 0.1 (continuous), 0.2 (dashed) and 0.3 (dot-dashed), for an input thermal state.}
\label{fig8}
\end{figure}

We plot the $Q$ parameter against $t$ for different values of $\bar{n}$ (thermal field) in figure~\ref{fig8}. $Q$ exhibits sub-Poissonian character for the input thermal state, and increases its negativity as $\bar{n}$ increases. But at $t=1$, the $Q$ parameter suddenly changes its characteristic to mark super-Poissonian statistics. Positive $Q$ parameter values are also observed in the negative range of $t$. We emphasize that though the SUP operated thermal field has no negative Wigner function, it displays sub-Poissonian property. Hence, Mandel's $Q$ parameter is a good indicator of the nonclassicality of the SUP operated thermal states.

The nonclassical nature of SOTS can also be analyzed by studying the squeezing parameter, which is calculated in terms of the average photon number $\bar{n}$,
\begin{equation}
S_{\mathrm{SOTS}} = 2N_2^{-1}\bar{n}\left[2\bar{n}(5s+3t)(s+t)+(2s+t)^2\right].
\end{equation}
\begin{figure}[ht]
\centering
\includegraphics[width=5cm,height=8cm,angle=-90]{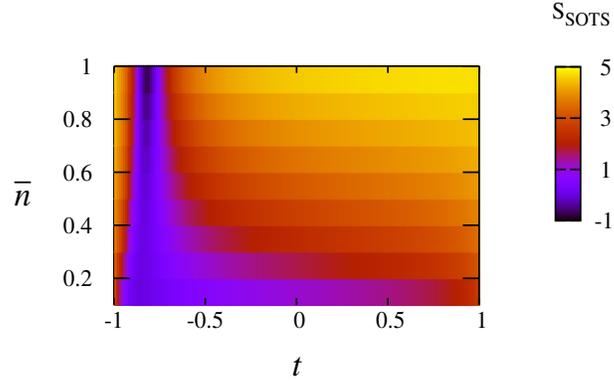}
\caption{Contour plot for $S_{\mathrm{SOTS}}$ as a function of $t$ and $\bar{n}$, for an input thermal state.}
\label{fig9}
\end{figure}

From figure~\ref{fig9}, it is clear that the squeezing parameter $S_{\mathrm{SOTS}}$ goes negative. Hence, the SUP operator can introduce the squeezing property into the input thermal state character.

\begin{figure}[ht]
\centering
\includegraphics[width=8cm]{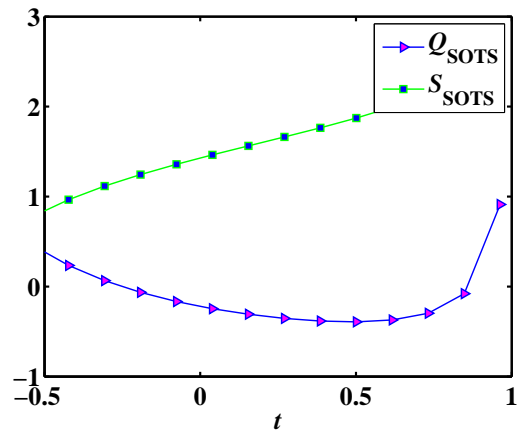}
\caption{A comparison of the different nonclassical indicators for SOTS as a function of the scalar parameter $t$ with $\bar{n}=0.2$.}
\label{com_thm}
\end{figure}

Hence we observe here that the Wigner function fails to indicate any nonclassicality for SOTS. Mandel's $Q$ parameter goes negative for some values of the control $t$. Unlike the input coherent state, SOTS shows squeezing property, and hence $S_{\mathrm{SOTS}}$ is a good measure. A comparison of the nonclassical indicators is shown in figure~\ref{com_thm}. The reason for the success of indicators such as the $Q$ parameter and the squeezing parameter $S$ over the non-negative Wigner function could be the negative values of normally ordered observables which are not realized by the Wigner function \cite{semenov}. As shown by the $F$-parametrized distribution function (figure~\ref{fig7.1}), such observables are determined by the Glauber-Sudarshan $P$ function. There are other examples of generated states that have non-negative Wigner functions and yet exhibit nonclassical features indicated by the $Q$ parameter \cite{cessa,janszky96}.

Non-Gaussian states are an important tool in quantum information processing and allow for applications of quantum algorithms which cannot be done using nonclassical Gaussian states. Nonclassical states with non-Gaussian Wigner function, as generated in the operation, have been used in the design for an efficient and universal quantum computation device \cite{Lloyd}. Non-Gaussian states have applications also in other quantum tasks \cite{Science,CC}.

\section{Conclusions}
\label{sec5}

The generation and manipulation of nonclassical field states is of great importance from the perspective of quantum tasks and information processing.
We have shown here that in continuous variable systems, a nonclassical state can be generated from an input classical state by the use of a general superposition of two product operations of the type $s(\hat{a}\hat{a}^\dag)+t(\hat{a}^\dag \hat{a}) = s + (s+t)\hat{a}^\dag \hat{a}$. The nonclassical property has been analyzed using the phase-space distribution of the generated state, its photon statistics and also the quadrature squeezing parameter.

The Wigner function is used to study the phase-space properties. The negativity of the Wigner distribution is a sufficient condition for the nonclassicality, and we have checked the area of negativity for different values of the scalar parameter $t$. For input coherent states, there is a distinct negativity in the Wigner function of the SUP operated output state. The negative volume for the SUP operated coherent state is shown to decrease with increasing $t$. Hence the negativity of the output state can be manipulated using the control parameter. For input thermal states,
the Wigner distribution is positive for all values of $t$. However, 
the SUP operated thermal state is non-Gaussian in nature. Thus the SUP operation can be used to generate non-Gaussian or nonclassical state for quantum operations on classical coherent or thermal fields. For the input coherent and thermal fields, we have checked for the negativity of the $F$-parametrized quasiprobability function as the operator ordering parameter $F$ is varied.

The photon statistics of the output state is another important indicator of nonclassicality. Field states with the classically most random photon distribution has a Poissonian character, in which the variance of the distribution is equal to the mean. Hence, any field with a sub-Poissonian photon distribution is essentially nonclassical. The sub-Poissonian statistics can be quantified using Mandel's $Q$ parameter. We have observed that the $Q$ parameter is negative (sub-Poissonian) for SUP operated coherent states for most values of the scalar parameter $t$ and is positive for values close to $t=$ 1. For SUP operated thermal states, the $Q$ parameter exhibits both super- and sub-Poissonian statistics based on the values of the control parameter $t$. Hence, the photon statistics of the output states can be controlled using $t$, which in turn generates the necessary nonclassicality.

Another important indicator of nonclassicality is the squeezing property of the output field state. The SOCS does not but the SOTS does exhibit squeezing property and hence the nonclassicality of the SOTS is well-described by the squeezing property.

The importance of generating nonclassical states using SUP operations is the fact that such operations can be realized experimentally using photon addition ($\hat{a}^\dag$) and subtraction ($\hat{a}$) properties of the input continuous variable states of the input field. The experimental generation of photon added and subtracted states in the laboratories using parametric down converters and controlled beam-splitters makes the physical realization of SUP operations.
Hence one can generate nonclassical states from classical distributions that can be suitably manipulated for specific quantum tasks with specific negativity of phase space distribution or sub-Poissonian statistics. It can also generate requisite non-Gaussianity from Gaussian states. These output states can thus have a wide range of useful applications.

From the perspective of quantum information applications in continuous variable regime, the generated output states can be used for various tasks that can be realized and measured in the laboratory. As mentioned earlier, Gaussian nonclassical states prove to be more useful than discrete quantum states in practical applications of various information protocols. On the other hand, states with non-Gaussian Wigner function can be used in the conceptual design of universal quantum computers and to implement quantum algorithms.

\ack
AC thanks National Board of Higher Mathematics, DAE, Government of India, and HSD thanks University Grants Commission, India, for financial support.


\begin{thebibliography}{99}
\bibitem{note} Detection of photons simultaneously at P2 and P3 along with P1 does not generate the desired product operation. It instead produces an operation $aa^\dag a$ which is not of interest for our study. We discard such operations.
\end{thebibliography}
\end{document}